\documentclass{iopart}

\usepackage{harvard}
\usepackage{latexsym}
\usepackage[dvips]{graphicx}
\usepackage{psfrag}

\def\be{\begin{equation}}
\def\bea{\begin{eqnarray}}
\def\bma{\begin{mathletters}}
\def\ee{\end{equation}}
\def\eea{\end{eqnarray}}
\def\ema{\end{mathletters}}
\newcommand{\beq}{\begin{equation}}
\newcommand{\beqa}{\begin{eqnarray}}
\newcommand{\eeq}{\end{equation}}
\newcommand{\eeqa}{\end{eqnarray}}
\newcommand{\nbeqa}{\begin{eqnarray*}}
\newcommand{\neeqa}{\end{eqnarray*}}
\newcommand{\nc}{\newcommand}
\nc{\rnc}{\renewcommand} \nc{\ket}[1]{\left | \, #1 \right
\rangle} \nc{\bra}[1]{\left \langle #1 \, \right |}
\nc{\proj}[1]{\ket{#1}\bra{#1}} \rnc{\vec}{\mathbf}
\nc{\braket}[2]{\langle\, #1\,|\,#2\,\rangle}
\nc{\half}{\frac{1}{2}}
\nc{\vfigure}[3]{
\begin{figure}[th]
\centerline{\psfig{file=figures/#1.eps,width=#2}}
\vspace*{8pt}
\caption{#3}
\end{figure}}
\nc{\vpstexfigure}[3]{
\vfigure{#1}{#2}{#3}}
\nc{\prj}{\mathcal{P}} \nc{\hilb}{\mathcal{H}}
\nc{\pth}{\mathcal{C}} \nc{\inprod}[2]{\braket{#1}{#2}}
\nc{\upket}{\ket{\uparrow}} \nc{\downket}{\ket{\downarrow}}
\nc{\upbra}{\bra{\uparrow}} \nc{\downbra}{\bra{\downarrow}}

\def\CC{{\rm\kern.24em \vrule width.04em height1.46ex depth-.07ex
\kern-.30em C}}
\def\P{{\rm I\kern-.25em P}}
\def\N{{\rm I\kern-.25em N}}
\def\RR{{\rm
         \vrule width.04em height1.58ex depth-.0ex
         \kern-.04em R}}
\def\id{{\rm 1\kern-.26em l}}
\def\ZZ{{\sf Z\kern-.44em Z}}

\def\i{{\rm i}\,}

\def\trace{{\rm tr}\;}

\newenvironment{eqblock}[2]{\beq\label{#2}\begin{array}{#1}}{\end{array}
                                \eeq}
\newenvironment{neqblock}[1]{\[\begin{array}{#1}}{\end{array}\]}

\newcommand{\beqb}{\begin{eqblock}}
\newcommand{\eeqb}{\end{eqblock}} 
\newcommand{\nbeqb}{\begin{neqblock}}

\newcommand{\neeqb}{\end{neqblock}}

\newcommand{\expect}[1]{\left\langle #1 \right\rangle}

\begin{document}

\title{Entanglement and magnetic order}
\author{Luigi Amico}
\address{MATIS-CNR-INFM $\&$ Dipartimento di Metodologie Fisiche e
    		Chimiche (DMFCI), viale A. Doria 6, 95125 Catania, Italy}
\author{Rosario Fazio}
\address{NEST, Scuola Normale Superiore  $\&$ CNR-INFM, Piazza dei Cavalieri 7, I-56126 Pisa, Italy\\
		and\\
		Center for Quantum Technology, National University of Singapore, 117542 Singapore, Singapore}


\begin{abstract}
In recent years quantum statistical mechanics have benefited of cultural interchanges with 
quantum information science. There is a bulk of evidence that quantifying the entanglement  
allows a fine analysis of many relevant properties of many-body quantum systems. 
Here we review the relation between 
entanglement and the various type of magnetic order occurring in interacting 
spin systems. 
\end{abstract}


\tableofcontents

\section{Introduction}
\label{intro}

In one of the most influential papers  for the foundation  of quantum physics~\cite{Schroedinger1935}, entanglement was recognized as  
not "{\it one}  but rather {\it the} characteristic trait of 
quantum mechanics".
Although such kind of non-local correlations were thoroughly explored for the analysis of the conceptual fundaments of 
quantum mechanics~\cite{Bell87,Peres93},
the  interest in understanding the properties of entangled states has received an impressive boost with the advent of quantum 
information. In fact it was understood that non-local correlations are responsible for the enhanced efficiency of quantum protocols~\cite{nielsen}.
 In some other cases,  quantum teleportation just to mention an important example, quantum correlations are a  necessary ingredient.  
Entanglement  can thus be considered a resource for quantum information processing.  The need for defining the amount of 
resources required to implement a given protocol has lead to a fertile line of research in quantum information aiming at the 
quantification of entanglement~\cite{horodeckirev}. To this end, necessary criteria for any entanglement measure to be fulfilled have 
been elaborated and lead to the notion of an entanglement monotone~\cite{MONOTONES}. 
There is a fairly clear scenario in the case of  bipartite systems. The multipartite counterpart is much more intricate.

The large body of knowledge developed in the characterization of entanglement has found important applications in areas that, less than a 
decade ago, were quite distant from quantum information. This is the case of quantum statistical mechanics.  Traditionally the characterization 
of many-body  systems has been carried on through the study of different physical quantities  (as the magnetization in magnetic systems)  and
their correlations. Very little attention was paid  to the structure of their quantum state and in particular to its amount of entanglement.  
The research at the border between these areas is rapidly evolving and has lead to numerous interesting results. The promise of this new 
interdisciplinary field  is twofold: firstly, to improve our understanding of strongly correlated systems, beyond the current state of the art; and 
secondly, to control and  manipulate quantum correlations  to our convenience, ultimately with the hope of making an impact on the 
computation schemes  to solve  'hard  problems' in computer science. 

Methods developed in quantum information have proven to be extremely  useful in the analysis of the state of many-body systems.  
At the same time experience built up over the years in condensed matter is helping in finding new protocols for quantum computation and 
communication.  The cross-fertilization of quantum information with statistical mechanics should not, however, come as a  surprise.  
Quantum computers are themselves  many-body systems, the main difference from traditional solid-state systems is that  a quantum computer 
can be controlled and operate under non-equilibrium conditions.  It is therefore  natural to profit of the best instruments developed so far in the 
two disciplines for the understanding of quantum complex systems, being quantum computers or condensed matter systems.  
In this review we deal with a particular aspect of this research area: We will focus on interacting spin systems on a lattice and describe the 
relations between  entanglement and magnetic ordering.  A more detailed discussion on these issues can be found in~\cite{amicorev,vlatkorev}.  

A complete characterization of entanglement in a many-body system is hopeless. The number of possible ways in which the 
system can be partitioned explodes on increasing  the number of elementary constituents (namely spins). Nevertheless
 judicious choices have made possible to highlight interesting properties of many-body correlations.  
Numerous contributions in this volume consider a bipartition in which  the system is divided into two distinct regions. 
If the total system is in a pure state, as when the  system is in its ground state,  then a measure of the entanglement  between 
the two regions is given by the von Neumann entropy associated to the reduced density matrix (of one of the two regions). 
This is not however the  only possible choice, our paper will briefly review other way to quantify entanglement whose 
properties contribute to our understanding of  many-body systems.

In the case of bipartite entanglement one can consider the quantum correlation between two given spins  after having traced out the rest of the system. In this case the entanglement between the two selected sites can be quantified by the concurrence~\cite{Wootters01}. 
The study of two-site entanglement, as we will briefly describe in the following of this paper,  allows to detect quite well the  presence of 
quantum phase transition in the phase diagram.  A different approach to entanglement in many-body systems arises from the  
quest to swap or transmute different types of multipartite entanglement into pairwise entanglement between two parties by 
means of generalized measures on the rest of the system. In a system  of interacting spins on a lattice this means to maximize 
the  entanglement between two spins by performing measurements on all the others. With this aim, the concept of localizable entanglement 
has been introduced in~\cite{verstraete04,popp05}.  

The structure of a many-body system's quantum state  is by far much richer than that captured by bipartite entanglement. In the multipartite 
case the grounds for quantitative predictions are less firm because of the exceptional difficulty of the problem. Nevertheless there are a
number of very interesting results already available. Among all we mention here the bounds which have been derived on the 
ground state energy~\cite{guehne05} which allow to discriminate among different n-particle quantum correlations. 
The idea of deriving bounds for macroscopic quantities (as for example the ground state energy) is a quite powerful methods to 
characterize entanglement in many-body system and is related to the concept of entanglement witness. As an interesting connection 
between statistical mechanics and quantum information it turns out that in many cases thermodynamic quantities as the magnetization 
or the susceptibility (in the case of interacting spin model) behave as entanglement witness~\cite{Toth05a,marcin}  thus providing a 
way to detect entanglement experimentally. Finally we would like to mention that the natural dynamics of condensed matter systems may 
be important to detect bound entangled states~\cite{patane.bound,acinbound} which are, in several cases, difficult to be realized artificially. 

Topic of this short review is to complement the different papers in this volume by describing various measure of entanglement, 
 other then the block entropy, which were used to characterize the equilibrium and dynamical properties of entanglement in 
spin systems (see also the article by J. I. Latorre and A. Riera in the same volume). Interacting spin models~\cite{spin-textbook,Schollwock} 
 provide  a paradigm to describe a wide range of many-body systems.  They account for the effective interactions in a variety of very 
different physical contexts ranging from high energy to nuclear physics. In condensed matter beside describing the properties of 
magnetic compounds, they capture several aspects of high-temperature superconductors, quantum Hall systems, heavy fermions,
just to mention few important examples.  Of course  interacting spins are central to quantum information processing~\cite{nielsen}.

The paper  is organized as follows. In the next Section we list the spin models that will be considered thereafter. Then 
we give a brief overview of the entanglement measures that are currently used to analyze many-body systems. In the 
Section~\ref{pairgr} we discuss the main outcomes of the uses of this entanglement measure applied to the spin models.  
The conclusions and a possible outlook will be presented in Section~\ref{concl}.

\section{Model systems}
\label{mod}

A large class of relevant models of interacting spin on a $d$-dimensional lattice can be described by the Hamiltonian
\begin{equation}
\label{general-spin}
  			{\cal {H}} {=}  \frac{1}{2}\sum_{i,j} 
                             \left[ J_x^{(ij)} S_i^x S_{j}^x +
                             J_y^{(ij)} S_i^y S_{j}^y 
                            	 + J_z^{(ij)} S_i^z S_{j}^z \right]- h^z \sum_{i} S_i^z \; .
\end{equation}
In Eq.(\ref{general-spin})  $S_i^\alpha$ ($\alpha=x,y,z$) are spin-$1/2$ operators defined on the $i$-th site of a $d$-dimensional lattice.  
The ground state of Eq.(\ref{general-spin}) is, in general, a highly entangled state. Nonetheless it exists  a point  where the 
ground state is indeed a classical state, factorized in the direct space~\cite{Kurmann82,Firenze05,giampaolofactor}. Such phenomenon occurs  at a precise value of the  field  $h_{\rm f}$ that was 
obtained in any dimensional bipartite lattice and for finite range exchange interaction~\cite{giampaolofactor} and even in presence of 
frustration~\cite{giampaolo_frustration}. Despite the similarities with the saturation phenomenon occuring in ferromagnets in external 
magnetic fields, it was proved that the factorization of the ground state is due to a  fine tuning of the control parameters, {\it within the 
magnetically ordered phase}. Interestingly enough, the results obtained so far indicate that the factorization point is a precursor of the 
quantum phase transition. We also mention  that analysis of the ground state for finite size systems demonstrated that the factorization 
can be viewed as transition between ground states of different parity\cite{rossignoli08,giorgi}. 

For nearest neighbors, the previous Hamiltonian defines the $XYZ$ anisotropic  Heisenberg model. In this case the exchange couplings  
are  commonly parameterized as $J_x=J(1+\gamma)$, $J_y=J(1-\gamma)$, and $J_z=2J \Delta$. A positive (negative) exchange 
coupling $J$ favors antiferromagnetic (ferromagnetic). In one dimension the model defined by the Hamiltonian~\ref{general-spin} is 
exactly solvable in several important cases. This  is particularly interesting in the analysis of entanglement because quantum effects  are 
particularly pronounced in low  dimensions.  

\begin{figure}
\begin{center}
\includegraphics[scale=0.5]{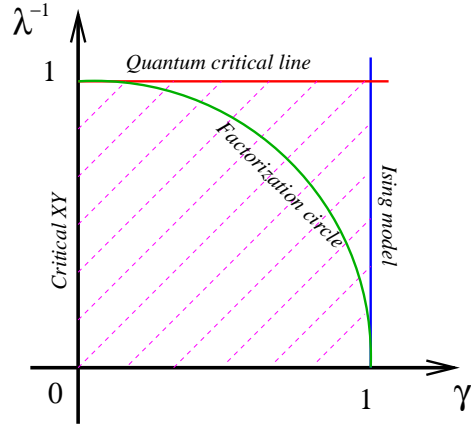} 
\end{center}
\caption{The zero temperature phase diagram of the one dimensional anisotropic $XY$ model in transverse field. Along the quantum critical line the model identifies the Ising 
		universality class with indices $z=\nu=1$; in the hatched 
area (with $\gamma>0$) the system display long range order in the $x-y$ spin components.
The critical $XY$ regime coicide with that one of the XXZ model for $\Delta=0$.
The factorization of the ground state occurs along the circle $\lambda^{-1}=\sqrt{1-\gamma^2}$.}
\label{xydiagr}
\end{figure}

Whenever $\Delta=0$ the (quantum anisotropic $XY$) Hamiltonian can be diagonalized by  first applying  the Jordan-Wigner transformation 
and then performing a Bogoliubov transformation~\cite{LIEB,PFEUTY,McCOY}. 
The  quantum Ising  model corresponds to  $\gamma =1$ while the (isotropic) $XX$-model is recovered for $\gamma = 0$. In the  isotropic case 
the model possesses an additional symmetry resulting in the conservation of the total magnetization along the $z$-axis. The properties 
of the Hamiltonian are governed by  the dimensionless coupling constant $\lambda=J/2h$. The phase diagram is sketched in Fig. \ref{xydiagr}
In the interval  $0<\gamma\le 1$ the system undergoes a second order quantum phase transition at the critical value $\lambda_c=1$. The 
order parameter is the magnetization in $x$-direction,  $\langle S^x\rangle $, different from zero for $\lambda >1$. In the phase with broken 
symmetry the ground state has a two-fold degeneracy reflecting a global phase flip symmetry of the system. The magnetization along the 
$z$-direction, $\langle S^z\rangle $, is different from zero for any value of $\lambda$, but is singular behavior in its first derivative at 
the transition point.   In the whole interval $0<\gamma\le 1$ the transition belongs to the Ising universality class. For $\gamma=0$ the quantum 
phase transition is of the  Berezinskii-Kosterlitz-Thouless type.  In several cases  the evaluation of entanglement requires the determination of 
the average magnetization $ M^{\alpha}_l(t) = \langle \psi | S_l^\alpha(t) |\psi  \rangle$ and of the equal-time correlation functions
$ g^{\alpha\beta}_{lm}(t)  = \langle \psi | S_l^\alpha(t) S_m^\beta(t)| \psi  \rangle$. These correlators  have been calculated for this class of models 
in the case of thermal equilibrium~\cite{LIEB,PFEUTY,McCOY}. These can be recast in the form of Toeplitz determinants in equilibrium case  
and  can be expressed as a  sum of Pfaffians in certain non-equilibrium situations~\cite{Amico-Osterloh}.

\begin{figure}
\begin{center}
\includegraphics[scale=0.5]{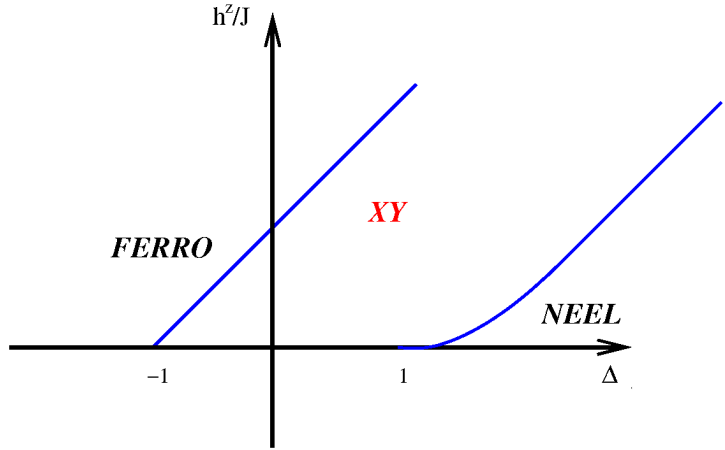} \end{center}
\caption{Zero temperature phase diagram of the spin $1/2$ XXZ model in one dimension. The $XY$ phase is characterized by power law 
		decay of the $xy$ correlations. The Neel and the XY phases are separated by a line of second order phase transitions; 
		at $\Delta=1$ the transition is of Berezinskii-Kosterlitz-Thouless. The ferromagnetic and $XY$ phases are separated by first 
		order phase transitions due to simple level crossings; the onset to the ferromagnetic phase occurs through  the saturation phenomenon.}
\label{xxzdiagr}
\end{figure}

In the case in which $\gamma=0$ and for any value of $\Delta$ the model is referred to as the $XXZ$ model. The two isotropic points $\Delta=1$ 
and $\Delta=-1$ describe the antiferromagnetic and ferromagnetic chains respectively (see the phase diagram in Fig.\ref{xxzdiagr}). The 
one-dimensional $XXZ$ Heisenberg model  can be solved exactly by Bethe Ansatz technique (see e.g.~\cite{Takahashibook99}) and the  correlation 
functions can be expressed in terms of certain determinants (see~\cite{Korepin-book} for a review).  Correlation functions, especially for intermediate 
distances, are in general  difficult to evaluate, although important steps in this direction have been made~\cite{Mallet,Korepin-inverse,ghoemann09}. 
The zero temperature phase diagram of the $XXZ$ model in zero magnetic field  shows a gapless phase in the interval $-1\le \Delta < 1$ with power 
law decaying correlation functions~\cite{Mikeska_xx,Tonegawa81}.  Outside this interval the excitations are gapped. The two phases are separated by a 
Berezinskii-Kosterlitz-Thouless  phase transition at $\Delta=1$ while at $\Delta=-1$  the transition is of the first order.  In the presence of the  external 
magnetic field a finite  energy gap appears in the spectrum. The universality class of the transition is not affected, as a result of the conservation of 
the total spin-$z$  component~\cite{Takahashibook99}.  

Another interesting case of the Hamiltonian in Eq.(\ref{general-spin}) is when  each spin interacts with all the other 
spins in the system with the same coupling strength 
$$ 
 {\cal {H}} {=}-{{J}\over{2}}   \sum_{ij} \left[  S_i^x S_{j}^x +\gamma S_i^y S_{j}^y \right]-  \sum_{i} \vec{h}_i \cdot \vec{S}_i \; .
 $$
For site-independent magnetic field $h_i^\alpha=h^\alpha \, \forall i\,, \; \alpha=x,y,z$, 
this model is known as the   Lipkin-Meshkov-Glick (LMG) model~\cite{LIPKIN1,LIPKIN2,LIPKIN3}.
In this case the dynamics of the system can be described in terms of a collective 
spin $S_\alpha=\sum_jS^\alpha_j$. At  $J^2\gamma=4 h^z$ the Hamiltonian manifests a supersimmetry~\cite{Unanyan-SUSY}.
The phase diagram depends on the parameter
$\lambda=J/2h^z$. For a ferromagnetic coupling ($J>0$) and $h^x=h^y=0$ the system undergoes a second 
order  quantum phase transition at $\lambda_c=1$, characterized by mean 
field critical indices~\cite{PFEUTY-LONG}. 
For $h^y=0$, $h^z < 1$ and $\gamma=0$ the model exhibits a first order transition at 
$h^x=0$~\cite{arias} while for an antiferromagnetic coupling and $h^y=0$
a first order phase transition at $h^z=0$ occurs  for  any $\gamma$'s. 
The model Hamiltonian defined above embraces an important class 
of interacting fermion systems with pairing force interaction (like the BCS model). Both the LMG and the BCS type  models 
can be solved exactly by Bethe  Ansatz~\cite{Richardson,Richardson-Sherman}.

We end this very brief overview with spin-$1$ systems which where originally considered to study the quantum 
dynamics of magnetic solitons in antiferromagnets  with single ion anisotropy~\cite{Mikeska}. 
In one dimension, half-integer and integer spin chains have very different 
properties~\cite{HALDANE-CONJ1,HALDANE-CONJ2}. We will see that the typical ground state of such models displays 
characteristic features in its entanglement content.  Long range order that is established
in the ground state of systems with half-integer spin~\cite{LIEB}, 
may be  washed out for integer spins. In this latter case, the system has a gap in the 
excitation spectrum. A paradigm model of interacting 
spin-$1$ systems is
\begin{equation}
H=\sum_{i=0}^{N}\vec{S}_i \cdot \vec{S}_{i+1}+\beta (\vec{S}_i \cdot \vec{S}_{i+1})^2
\label{haldane-phase}
\end{equation} 
The lack of long range order  arises because of the presence of
zero as an eigenvalue of $S^z_i$; the corresponding 
eigenstates represent a spin configuration that can move freely in the chain, 
ultimately disordering the ground state of the system, experiencing a gap 
with the lowest energy excitation~\cite{Mikeska,Gomez-Santos}.
The so called string order parameter was proposed to 
capture the resulting 'floating' N\'eel order, made of alternating spins 
$|\uparrow\rangle$,  $|\downarrow \rangle$ with strings of $|0\rangle$'s
in between~\cite{denijs} 
\begin{equation}
                 \displaystyle{O_{string}^\alpha=
                  \lim_{R\rightarrow \infty}
                   \langle S^\alpha_i (\prod_{k=i+1}^{i+R-1}
                  e^{i\pi  S^\alpha_k})S^\alpha_{i+R}\rangle } \;\; .
\label{string}
\end{equation}
The ground state of physical systems described by Hamiltonians of the form of 
Eq.(\ref{haldane-phase}) has been studied in great details~\cite{Schollwock}. 
Different phase transitions have been found between antiferromagnetic phases, 
Haldane phases, and a phase characterized by a large density of vanishing 
weights ($S^z_i=0$) along the chain. 
Some features of the phenomenology 
leading to the  destruction of the antiferromagnetic  order  can be put on a 
firm ground for  $\beta=1/3$ (AKLT model), where the ground state of the Hamiltonian in 
Eq.(\ref{haldane-phase}) is known exactly~\cite{AFFLECK}. In this case it was 
proved that the ground state is constituted  by a sea of nearest neighbor  
valence bond states, separated from the first excitation by 
a finite gap with exponentially decaying correlation functions.

\section{Entanglement measures}
\label{meas}
The study of quantum correlations in many-body systems depends heavily on the impressive progress that has been 
achieved in the theory of entanglement quantification. The new ingredient that makes the many-body case very 
appealing is the rich variety of ways in which the system can be 
partitioned into subsystems.  
Comprehensive overviews 
of  entanglement measures can be found in~\cite{bengtsson,Bruss01,Th-Eisert,horodeckirev,Plenio98,Plenio07,Vedral02,Wootters01}, it 
is however convenient to recall some of the entanglement measures that are routinely used to characterize many-body systems. 
Important requirements for an entanglement measure are that it should be invariant under local unitary operations; it should 
be continuous and, furthermore,  additive when several identical copies are considered. 

Most of the work done on entanglement in many-body systems deals with the bipartite case. 
A pure bipartite state is not entangled if and only if it can be written as a tensor product of pure states of the parts.
It can be demonstrated that 
reduced density matrices can be decomposed by exploiting the Schimdt decomposition: 
$
\rho_{B/A} = \sum_{i} \alpha_{i}^2\ket{\psi_{B/A,i}}
\bra{\psi_{B/A,i}}  
$.
Since only product states lead to pure reduced density matrices, a measure for their mixedness points a 
way towards quantifying entanglement. One can thus use a suitable function of the $\alpha_i$ given by the 
Schmidt decomposition to quantify the entanglement. Remarkably enough the 
von Neumann entropy
$
S(\rho_{B/A}) =\sum_i \alpha_i^2 \log(\alpha_i^2) \; ,
$ 
can {\it quantify} the entanglement encoded in $\rho_{B/A}$. We point out that an infinite class of entanglement measures can be constructed for pure states. 
Infact by tracing out one of two qubits in the  state, the corresponding reduced density matrix $\rho_A$ contains only a single 
independent  parameter: its eigenvalue $\leq 1/2$.
This implies that each monotonic function $[0,1/2]\mapsto [0,1]$ of this eigenvalue
can be used as an entanglement measure.  A relevant example is the (one-) tangle~\cite{Coffman00}
$\tau_1[\rho_A] = 4 {\rm det} \rho_A$.
By expressing $\rho_A$ in terms of spin expectation values, it follows that 
$
\tau_1[\rho_A]=\frac{1}{4}-
(\expect{S^x}^2+\expect{S^y}^2+\expect{S^z}^2) 
$
where $\expect{S^{\alpha}} = tr_A(\rho_A S^{\alpha})$ and $S^{\alpha}=\frac{1}{2} \sigma^\alpha$, 
$\sigma^\alpha$ $\{\alpha=x,y,z\}$ being the Pauli matrices,
For a pure state  of two qubits it can be shown that 
$\tau_1$ is equivalent to the
concurrence $C$~\cite{Hill97,Wootters98} for pure states of two qubits.
The von Neumann entropy can be expressed as a function of the (one-) tangle
$\displaystyle{
S[\rho_A]=h[(1+\sqrt{1-\tau_1[\rho_A]})/2]}
$
where
$h(x)=: -x \log_2 x - (1-x) \log_2 (1-x)$ is the binary entropy.

Subsystems of a many-body (pure) state will generally be in a mixed state.
In this case different ways of quantifying entanglement can be realized.
Three important representatives are the entanglement cost $E_C$,
the distillable entanglement $E_D$  (both defined in Ref.~\cite{Bennett96}) and the 
entanglement of formation $E_F$~\cite{BennettDiVincenzo96}. In the following we concentrate on 
the entanglement of formation. 
The conceptual difficulty behind its calculation  lies
in the infinite number of possible decompositions of a density matrix.
Therefore, even knowing how to quantify bipartite  entanglement in pure states, we cannot simply apply this knowledge 
to mixed states in terms of an average over the mixtures of 
pure state entanglement. 
It turns out that the correct procedure is to take the minimum over all possible decompositions.
This conclusion can be drawn from the requirement that entanglement 
must not increase on average by means of local operations including 
classical communication.  The entanglement of formation of a state ${\rho}$ is 
therefore defined as 
\begin{eqnarray}
E_{F}(\rho):= \min \sum_j p_j S(\rho_{A,j}) \; , \label{ef}
\end{eqnarray}
where the minimum is taken over all realizations of the state 
$\rho_{AB} = \sum_j p_j |{\psi_j}\rangle\langle{\psi_j}|$, and $S(\rho_{A,j})$ 
is the von Neumann entropy of the reduced density matrix 
$\rho_{A,j}:=\trace_B \ket{\psi_j}\bra{\psi_j}$.
For systems of two qubits, an analytic expression for $E_F$ does exist
and it is given by
\begin{equation}
\label{EoF}
E_F (\rho) = -\sum_{\sigma = \pm}\frac{\sqrt{1+\sigma 
C^2(\rho)}}{2}\ln\frac{\sqrt{1+\sigma C^2(\rho)}}{2}
\end{equation}
where $C(\rho)$ is the the so called concurrence~\cite{Wootters98,Wootters01} defined as
\begin{equation}
  \label{eq:concurrence}
  C = \mbox{max}\{\lambda_1-\lambda_2-\lambda_3-\lambda_4,0\}\; .
\end{equation}
where $\lambda_1^2\geq \dots \geq \lambda_4^2$ are the eigenvalues of 
$
 R\equiv\sqrt{\rho}\tilde{\rho}\sqrt{\rho} =\sqrt{\rho}(\sigma^y\otimes\sigma^y)
\rho^*(\sigma^y\otimes\sigma^y)\sqrt{\rho}
$ ($*$ indicates the complex conjugation).
As the entanglement of formation is a monotonous function of the concurrence,  also $C$ itself or its square $\tau_2$ - called 
also the 2-tangle - can be used as entanglement measures.  The concurrence $C$ and the tangle $\tau_1$ both range 
from $0$ (no entanglement) to $1$.
By virtue of (\ref{eq:concurrence}), the concurrence in a spin-1/2 chain can be computed
in terms of the two-point spin correlation functions.  
As an example (that is relevant for the present article) we consider a case where the model has a 
parity symmetry, it is translational invariant and the Hamiltonian is real. 
In this case the concurrence reads
\begin{equation}
\label{C-of-corrs}
C_{ij}=2\max\left\{0,C_{ij}^I,C_{ij}^{II}\right\}\;.
\end{equation}
where
$
C^I_{ij}=|g_{ij}^{xx}+g_{ij}^{yy}|-\sqrt{\left(1/4+g^{zz}_{ij}\right)^2-M_z^2}
$
and 
$
C^{II}_{ij}=|g_{ij}^{xx}-g_{ij}^{yy}|+g_{ij}^{zz}-1/4
$, 
with $g_{ij}^{\alpha\alpha}=\langle S^\alpha_i S^\alpha_{j} \rangle$
and $M_z=\langle S^z \rangle$.
A state with dominant fidelity of parallel and anti-parallel Bell states is 
characterized by dominant $C^I$ and $C^{II}$, respectively. 
This was shown in~\cite{Fubini06}, 
where the concurrence was expressed in terms of the fully entangled 
fraction as defined in~\cite{BennettDiVincenzo96}.

The importance of the tangle and the concurrence is due to the monogamy of entanglement which 
was expressed in terms of an inequality in~\cite{Coffman00} for the case of three qubits. 
This inequality has been 
proved to hold also for n-qubits system~\cite{Osborne06}. In the case of 
many-qubits it reads 
$
        \sum_{j\neq i} C^2_{ij} \leq  \tau_{1,i} \; .
$
The so called residual tangle $\tau_{1,i}-\sum_{j\neq i} C^2_{ij}$, is 
a measure for multipartite entanglement
not stored in pairs of qubits only.

Another measure of entanglement we mention is the relative entropy 
of entanglement~\cite{RelativeEntropy}. It can be applied to any number of 
qubits in principle (or any dimension of the local Hilbert space).
It is formally defined as
$
E({\sigma}):= \min_{\rho \in {\cal D}}\,\,\, S(\sigma || \rho)
$,
where 
$S(\sigma || \rho) = \trace \sigma \left[\ln \sigma - \ln \rho \right]$
is the quantum relative entropy. This relative entropy of entanglement
quantifies the entanglement in $\sigma$ by its distance from the set ${\cal D}$
of separable states. 
The main difficulty in computing this measure is to find the 
disentangled state closest to $\rho$. This is in general a non trivial  
task, even for two qubits. In the presence of certain symmetries 
- which is the case for e.g. eigenstates of certain models - an analytical 
access is possible. In these cases, the relative entropy of entanglement  
becomes a very useful tool. The relative entropy reduces to the 
entanglement entropy in the case of pure bi-partite states; this also 
means that its, so called, {\it convex roof extension} \cite{Uhlmann97} coincides with the 
entanglement of formation, and is readily deduced from the 
concurrence~\cite{Wootters98}.

It is important to realize that not just the quantification of 
many-party entanglement is a difficult task; it is an open 
problem to tell in general, whether a state of $n$ parties is separable or not. 
It is therefore of great value to have a tool that is able to merely
certify if a certain state is entangled. An entanglement witness $W$ 
is a operator that is able to detect entanglement in a state. 
The basic idea is
that the expectation value of the witness $W$ for the state $\rho$
under consideration exceeds certain bounds only when $\rho$ is entangled. 
An expectation value of $W$ within this bound however
does not guarantee that the state is separable. Nonetheless, this is a very 
appealing method also from an experimental point of view, since it is 
sometimes possible to relate the presence of the entanglement to the 
measurement of few observables.
Simple geometric ideas help to explain the  witness 
operator $W$ at work. Let $\mathcal{T}$ be the set of all density matrices 
and let $\mathcal{E}$ and $\mathcal{D}$ be the subsets of entangled and 
separable states, respectively. 
The convexity of $\mathcal{D}$ is a key property for witnessing 
entanglement
The entanglement witness is then an operator 
defining a hyper-plane which separates a given entangled state from the set of 
separable states. The main scope of this geometric approach is then 
to optimize the witness operator~\cite{LewensteinOptWit} or to replace 
the hyper-plane by a curved manifold, tangent to the set of separable 
states~\cite{witguehne}.
We have the freedom to choose $W$ such that 
$
\trace(\rho_D W)\leq 0
$ 
for all disentangled states $\rho_D\in{\cal D}$.
Then,  
$
\trace(\rho W)>0
$ 
implies that $\rho$ is entangled.
Entanglement witnesses are a special case of a more general concept, 
namely that of positive maps. These are injective superoperators
on the subset of positive operators.
When we now think of super-operators acting non-trivially only on part
of the system (on operators that act non trivially only on a sub-Hilbert space),
then we may ask the question whether a positive map on the subspace is 
also positive when acting on the whole space. Maps that remain positive 
also on the extended space are called completely positive maps.
Positive but  not completely positive 
maps are important for entanglement theory. Indeed it can be shown~\cite{Horodecki96}
that  state $\rho_{AB}$ is entangled if and only if
a positive map $\Lambda$ exists (not completely positive) such that 
$
(\id_A \otimes \Lambda_B) \rho_{AB} < 0
$.
For a two dimensional local Hilbert space the situation simplifies
and in a
system of two qubits the lack of complete positivity in a positive
map is due to a partial transposition. This partial transposition 
clearly leads to a positive operator if the state is a tensor 
product of the parts. In fact, also the opposite is true: 
a state of two qubits $\rho_{AB}$ is separable if and only if 
$\rho_{AB}^{T_B} \geq 0$ that is, its partial transposition is positive. 
This is very simple to test and it is known as the Peres-Horodecki 
criterion~\cite{Peres96,Horodecki96}. The properties of entangled 
states under partial transposition lead to a measure of entanglement
known as the {\em negativity}. The negativity $N_{AB}$  
of a bipartite state is defined as the absolute value of the sum of 
the negative eigenvalues of $\rho_{AB}^{T_A}$. The 
{\em logarithmic negativity} is then defined as
\begin{equation}
     E_N  = \log_22 (2N_{AB}  + 1).
\end{equation}
For bipartite states of two qubits, $\rho_{AB}^{T_A}$ has at most one 
negative eigenvalue~\cite{Sanpera98}. For general multipartite and 
higher local dimension this is only a sufficient condition for the 
presence of entanglement.  There exist entangled states with a
positive partial transpose (PPT) known as bound entangled 
states~\cite{acinbe,Horodeckibe}. The existence of bound entangled states ultimately limits the possibility  
to exploit the  violation of Bell inequalities as a measure of entanglement. 

Let us briefly introduce the concept of bound entanglement. 
Such kind of entanglement can be recognized in terms of the so called distillation protocol, 
a non trivial procedure to optimize 
the extraction of Bell states from a mixture of entangled states~\cite{Bennett96-distill}.
 The natural question, then is: Can  any entangled state be  actually distilled? The answer is yes for bipartite and qubit 
 states~\cite{horodecki97}. For multipartite entangled states and higher dimensional local Hilbert spaces  a much more 
 complex scenario emerges. In these cases examples of entangled states have been provided that cannot be distilled 
 to  maximally entangled states  between the parties of the system,  not even with an asymptotically infinite supply of  
 copies of the state. Such a demoted form of entanglement was termed as bound entanglement.  
PPT entangled states were found first in~\cite{Horodeckibe}. The existence of bipartite (with higher dimensional local 
Hilbert spaces) NPT bound entanglement has not excluded yet. This question has important implication on the additivity 
property of the distillable entanglement: if affirmative, entangled states could be generated from a mixture of non-distillable 
states. Besides its speculative interest, it was demonstrated that bound entanglement  can be activated in several 
quantum information and teleportation tasks to 'restore' the singlet fidelity of a given state (see~\cite{horodeckirev}).

Multipartite bound entangled states exist that are not  fully  separable but contain entanglement between each of their parties 
that cannot be distilled. Given that violation of PPT is necessary for distillation, a feature of multipartite bound entanglement   is related to the 
'incomplete separability' of the state (see~\cite{DUR-CIRAC}).
A  tripartite system A-B-C, for example, is separable  with respect to the partition A|BC and B|AC and non-separable with respect to C|AB. 
The 'incomplete separability' is a sufficient condition for a state to have  bound entanglement  since the three qubits are entangled and 
no maximally entangled state can be created between any of the parties by LOCC.
 For example, no entanglement can be  distilled  between C and A because  no entanglement can be created with respect to the 
 partition A|BC by LOCC.
 In the nest section  the feature of incomplete separability will be exploited to detect  multipartite bound entanglement  in spin  systems.   
The nature of  correlations in bound entangled states  is peculiar  having  both quantum and classical features. Therefore, if it is true 
that all entangled states violates Bell inequality, the vice versa has not proven (see also \cite{Popescu95} and Gisin \cite{Gisin96}). 
Bipartite bound entangled states seem not violating Bell 
inequalities~\cite{Masanes06} (but some examples of multipartite bound entangled state violating Bell-type inequalities 
were found~\cite{DUR2}). For both the bipartite and multipartite entangled states there is a strong believe that necessary and sufficient 
condition for local realism is that the state satisfies the Peres criterium~\cite{Peres,Werner01,WernerWolf00}. 

As already mentioned, a classification of multipartite entanglement is still missing. Nevertheless there are several quantities 
serving as indicators for multipartite entanglement when the whole system is in a pure state.
The entropy of entanglement is an example for such a quantity and several  
works use multipartite measures constructed from and related to
it (see e.g.~\cite{Coffman00,Wallach,Viola03,Scott04,oliveira,Love06}).
These measures give  indication on a global correlation without discerning among the different 
entanglement classes encoded in the state of the system.
The geometric measure of entanglement quantifies the entanglement of a 
pure state through the minimal distance of the state from the set of 
pure product states~\cite{RelativeEntropy,wei03}
\begin{equation}
   E_{g}(\Psi) = - \log_2 \max_{\Phi}\mid \langle \Psi | \Phi \rangle \mid ^2
\label{weimeas}
\end{equation}
where the maximum is on all product states $\Phi$.  It is zero for separable states 
and rises up to unity for e.g. the maximally entangled n-particle GHZ states. 
The difficult task in its evaluation is the maximization 
over all possible separable states and of course the convex roof extension to mixed states. 
Despite these complications, a clever use of the symmetries of the problem 
renders this task accessible by substantially reducing the 
number of parameters over which the maximization has to be performed.
Another example for the collective measures of multipartite
are the measures introduced by Meyer and Wallach~\cite{Wallach} 
and by Barnum {\em et al}~\cite{Viola03,Barnum04}. 
In the case of qubit system the measure of
Meyer and Wallach is the average purity (which is the average one-tangle 
in~\cite{Coffman00}) of the state~\cite{Wallach,Brennen03,Barnum04}
\begin{equation}
  E_{gl} = 2 - \frac{2}{N} \sum_{j=1}^N \mbox{Tr} \rho_j^2 \;\; .
\label{GE}
\end{equation}
The notion of generalized entanglement introduced 
in~\cite{Viola03,Barnum04} relaxes the typically chosen 
partition into local subsystems in real space. For the state $| \psi \rangle$ it is defined as 
\begin{equation}
  P_{\cal A} = \mbox{Tr} \left\{ \left[ {\cal P}_{{\cal A}} | 
\psi \rangle \langle \psi | \right]^2 \right\} 
\end{equation}
where ${\cal P}_{\cal A}$ is the projection map 
$\rho \rightarrow {\cal P}_{\cal A}(\rho)$. 
If the set of observables is defined by the operator basis 
$\left\{ A_1, A_2,\dots, A_L\right\}$ then
$  P_{\cal A} = \sum_{i=1}^L \langle A_{i}\rangle ^2
$
from which the reduction to Eq.(\ref{GE}) in the case of all local observables 
is evident. This conceptually corresponds to a redefinition of locality as induced by
the distinguished observable set.
Finally we mention the approach pursued in~\cite{guehne05}
where different bounds on the average energy of a given system were obtained 
for different types of n-particle quantum correlated states.
A violation of these bounds then implies the presence of 
multipartite entanglement in the system.
The starting point of G\"uhne {\em et al.} is the notion of 
{\em n-separability} and {\em k-producibility} which admit to discriminate 
particular types of n-particle correlations present in the system. 
A pure state $\mid \psi \rangle$ of a quantum systems of N parties is 
said to be n-separable if it is possible to find a partition of 
the system for which $\mid \psi \rangle = |\phi_1 \rangle |\phi_2 
\rangle \cdots |\phi_n \rangle$.
A pure state $\mid \psi \rangle$ can be 
produced by k-party entanglement (i.e. it is k-producible) if we can write
$\mid \psi \rangle = |\phi_1 \rangle |\phi_2 \rangle \cdots |\phi_m \rangle$
where the $|\phi_i \rangle$ are states of maximally k parties; 
by definition $m \ge N/k$. 
It implies that it is sufficient to generate specific 
k-party entanglement to construct the desired state. 
Both these indicators for multipartite entanglement are collective,
since they are based on the property  of a given many particle
state to be factorized into smaller parts. k-separability and -producibility both
do not distinguish between different k-particle entanglement classes
(as e.g. the k-particle W-states and different k-particle 
graph states~\cite{hein04}, like the GHZ state).  Another approach is based on the observed  relation between entanglement 
measures and  $SL(2,\CC)$ invariant antilinear 
operators\cite{Uhlmann,OS04,OS05,AndreasJMP}. This allows certain sensitivity to different classes of multipartite
 entanglement(see also \cite{VerstraeteDMV02,Lamata07,Bastin09,Andreas.summary}).  

We close this section by reviewing how to swap or transmute different types of multipartite 
entanglement in a many body system
into pairwise entanglement between two parties by means of 
generalized measures on the rest of the system. In a system 
of interacting spins on a lattice one could then try to maximize the 
entanglement between two spins (at positions $i$ and $j$) by performing 
measurements on all the others. The system is then partitioned in three 
regions: the sites $i$, $j$  and the rest of the lattice.
This concentrated pairwise entanglement can then
be used  e.g. for quantum information processing.
A standard example is that of  a GHZ state 
$(1/\sqrt{2})(\ket{000}+\ket{111})$. 
After a projective measure in $x$-direction on one of the sites 
such a state is transformed into a Bell state.
The concept of {\it localizable entanglement} has been introduced 
in~\cite{verstraete04,popp05}. 
It is defined as the maximal amount of entanglement that can be 
localized, on average, by doing local measurements in the rest of the 
system. In the case of $N$ parties, the possible outcomes of the measurements 
on the remaining $N-2$ particles are pure states $|\psi_s \rangle$ 
with corresponding probabilities $p_s$. The localizable entanglement 
$E_{loc}$ on the sites $i$ and $j$ is defined as the maximum of 
the average entanglement over all possible outcome states $\ket{\psi_s}_{ij}$
\begin{equation}
E_{loc}(i,j) = \mbox{sup}_{\cal{E}} \sum_s p_s E(\ket{\psi_s}_{ij})
\end{equation}
where ${\cal E}$ is the set of all possible outcomes 
$(p_s, |\psi_s \rangle)$ of the measurements, and $E$ represents 
the chosen measure of entanglement of a pure state of two qubits 
(e.g. the concurrence). 
Although very difficult to compute, lower and upper bounds have 
been found which allow to deduce a number of non trivial properties of the state. 
An upper bound to the localizable entanglement is given by the 
entanglement of assistance~\cite{laustsen03} obtained from 
localizable entanglement when also global and joint measurements 
were allowed on the $N-2$ spins .
A lower bound of the localizable entanglement~\cite{verstraete04} is fixed 
by the maximal correlation function between the two parties.
	
\section{Entanglement and magnetic order}
\label{pairgr}

The entanglement present in the equilibrium (thermal or ground) state of a quantum system is very sensitive to the 
underlying collective behavior. This suggests, in the case of spin systems, 
to analyze the relation between entanglement 
and magnetic order.  We will discuss various aspects of this connection starting from the pairwise entanglement, we then 
proceed with the properties of multipartite  entanglement.  Most of the investigations available in the literature are  for 
one-dimensional systems where exact results are available, later on we will overview the status in the d-dimensional case.

The body of knowledge acquired so far makes  it evident  that entanglement in the ground state contains relevant 
information on zero temperature phase diagram of the system. We will highlight this relation in two paradigmatic 
cases when the spin system is either close to a quantum phase transitions or to  factorizing field.  We will mostly 
be concerned with a  $XYZ$ spin models in an external field.

\begin{figure}\begin{center}
\includegraphics[scale=0.3,angle=0]{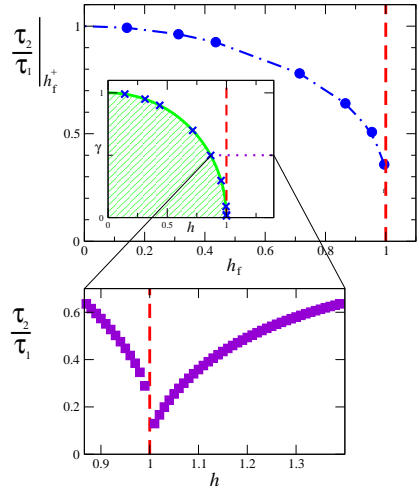}
\end{center}
\caption{One-tangle 
Entanglement ratio for the $XY$ model: 
the dashed line indicates the critical field.
Upper panel: Value of $\tau_2/\tau_1$ as $h\to h_{\rm f}^+$, as a function 
of $h_{\rm f}^{XY}=\sqrt{1-\gamma^2}$, for different values of $\gamma$; 
the Inset shows the "entanglement phase-diagram" in the $h-\gamma$ plane: 
the shaded (white) area corresponds to antiparallel (parallel) pairwise 
entanglement and the full line is the separable circle; the crosses 
indicate the position of the data shown in the main upper panel, while the 
dotted line corresponds to the $x$-axis used in the lower panel.
Lower panel: $\tau_2/\tau_1$ versus $h$, for $\gamma=0.5$.
[From\cite{Amico06.unpublished}]} 
\label{tau2tau1.fact}
\end{figure}

\paragraph{Range of pairwise entanglement}
As we discussed in Sec.~\ref{mod}, in an interacting spin system there exists a 
particular choice of the coupling constants and the external field for which the ground state is 
factorized~\cite{Kurmann82,giampaolofactor}, i.e. the entanglement vanishes exactly. Several works were devoted to 
the characterization of the entanglement close to the {\em factorizing point}.  It was demonstrated for one dimensional 
spin models in the class  $XXZ$ that the point at which the state of the system becomes separable marks an exchange 
of parallel and anti-parallel sector in the ground state concurrence\cite{Fubini06,Amico06}.  This change occurs through  
a global (long-range) reorganization of the state of the system.  The range $R$ of the concurrence (defined through 
the maximum distance between two sites over which it is non-zero) diverges.  For the $XY$ model it was found that 
this range is
\begin{equation}
R\propto \left (\ln \frac{1-\gamma}{1+\gamma}\right )^{-1} \ln |\lambda^{-1}-
\lambda_f^{-1}|^{-1}\; .
\end{equation}     
The existence of such  a divergence has been confirmed in other one dimensional systems both for 
short~\cite{Amico06,Firenze04,Roscilde.jltp} and long range interactions~\cite{DusVidPRL}.  This divergence  
suggests, as a consequence of the monogamy of the  entanglement~\cite{Coffman00,Osborne06},  that the role of 
pairwise entanglement is enhanced on  approaching the factorizing field~\cite{Firenze04,Roscilde.jltp,Firenze05}.
Indeed, for the Ising model (i.e. $\gamma=1$), one finds that in this region the ratio between the two-tangle and the 
one-tangle tends to one(Fig.\ref{tau2tau1.fact})~\cite{Amico06}. 

The diverging entanglement length is particularly intriguing in systems characterized by topological order. In Ref.\cite{wonmin} the
entanglement in the quasi-long-range ordered ground state of the one dimensional isotropic $XY$  model. 
Because of the presence of the characteristic edge states in  systems with non trivial 
topology, it  was found  that the quasi-long-range  order is traced  
by entangled states localized at the edges of the system. 
The range of the concurrence diverges also close to the saturation field 
for a spin system with  inverse-square interaction of the Haldane-Shastry
type\cite{Haldane-spin,Shastry}, interpolating in a sense between nearest-neighbor and fully connected graph interactions. 
In particular, in Ref.\cite{ent.Haldane-Shastry} it was shown that, 
in the absence of external magnetic field  the Haldane-Shastry spin 
system only displays nearest neighbor entanglement; 
while, by increasing the magnetic field, the bipartite entanglement between spins at greater distance increases, up to a situation where  the spin system saturates, entering a fully polarized phase described by a 
completely separable ground state. Given the special role of the inverse-square interaction in one dimensional fractional statistics\cite{Les-Houches}, this study is also relevant for the understanding of the interplay between the statistics and the entanglement. 

The range of concurrence does not diverge only at the factorizing field. There are one-dimensional spin systems where 
the pairwise entanglement has qualitative different features as a function of the distance between the sites. An example is 
the {\em long-distance entanglement} observed in~\cite{Venuti06b}.   Given a measure of entanglement $E(\rho_{ij})$, 
Campos Venuti {\em et al} showed that  it is possible that  $E(\rho_{ij}) \ne 0$ when $|i-j| \to \infty$ in the ground state. 
Long-distance entanglement can be realized in various one-dimensional models as in the  dimerized frustrated Heisenberg 
models or in the AKLT model.  For these two models the entanglement is highly non-uniform and it is mainly concentrated 
in the end-to-end pair of the chain.

\paragraph{Pairwise entanglement and quantum phase transitions}
A great number of papers have been devoted to the study  of entanglement close to quantum phase transition (QPT).  
Close to the quantum critical point the system is characterized by a diverging correlation length $\xi$ which is responsible 
for the singular behavior of different physical observables.  The behavior of correlation functions however is not necessarily 
related to the behavior of entanglement. It is worth to stress that the study of entanglement close to quantum critical points 
does not provide new understanding to the scaling theory of quantum phase transitions. Rather it may be useful in a deeper 
characterization of the ground state wave function of the many-body system undergoing a phase transition. In this respect 
it is important to explore, for instance, how the entanglement depends on the order of the transition, or what is the role 
of the range of the interaction to establish the entanglement in the ground state.  We start by considering exclusively the 
properties of  pairwise entanglement. 

\begin{figure}\centering
\includegraphics[width=8cm]{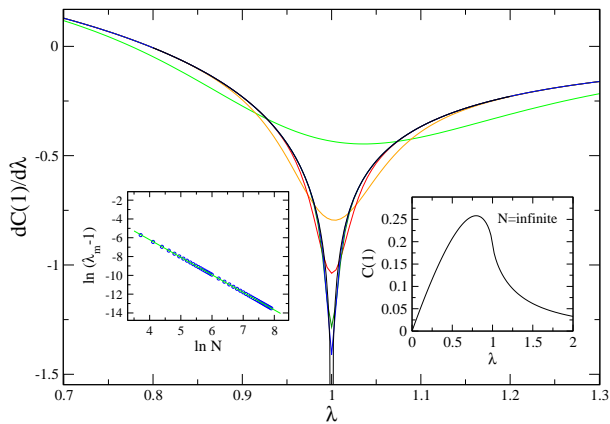}
\vspace*{0.3cm}
\caption{
The change in the ground state wave function in the critical region 
is analyzed considering the derivative of the nearest neighbor concurrence 
as a function of the reduced coupling strength. The  curves correspond to different 
lattice sizes. On increasing the system size, the minimum gets more pronounced. 
Also the position of the minimum changes and tends as  (see the left side inset) 
towards the critical point where for an infinite system a logarithmic divergence 
is present. The right hand side inset shows the behavior of the concurrence 
itself for an infinite system.[From\cite{OstNat}]}
\label{a-derivative-gs}
\end{figure}

Pairwise entanglement close to  quantum phase transitions was originally analyzed in~\cite{Osborne02,OstNat} for the 
Ising model in one dimension. Below we summarize their results in this specific case. The concurrence tends to zero for  
$\lambda\gg 1$ and $\lambda \ll 1$, the ground state  of the system is fully polarized along the $x$-axes ($z$-axes). 
Moreover the concurrence is zero unless the two sites  are at most next-nearest neighbors, we therefore discuss only 
the nearest neighbor concurrence $C(1)$. The concurrence itself is a smooth function of the coupling with a 
maximum close to the critical  point (but not related to any property of the phase transition).
The critical properties of the ground state are captured  by the derivatives of
the concurrence as a function of $\lambda$. The results  are shown in Fig.\ref{a-derivative-gs}.
In the thermodynamic limit  $\partial_\lambda C(1) $ 
diverges on approaching the critical value as 
\begin{equation}
\partial_\lambda C(1) \sim \frac{8}{3 \pi^2} \ln|\lambda-\lambda_c| \;\; .
\label{nn-concurrence}
\end{equation}      				
For finite system the precursors of the critical behavior 
can be analyzed by means of finite size scaling of the derivative of the concurrence. 
Similar results have been obtained for the $XY$ universality class~\cite{OstNat}.  Remarkably, although the 
concurrence describes short-range properties, nevertheless scaling behavior typical of continuous phase 
transition emerges.

Over the last years the properties of pairwise entanglement were intensively studied.  It was evidenced how it 
depends on the order of transition and on the universality class of the system. The bulk of results obtained so far 
can be summarized in a 'Ehrenfest classification scheme for entanglement', ultimately  arising because of  
the formal relation between the correlation functions and  the entanglement.  A way to put this observation on a 
quantitative ground is provided by a generalized Hohenberg-Kohn theorem~\cite{Wu06}.  Accordingly, the ground 
state energy can be considered as a unique function of the expectation values of certain observables. These, in turn, 
can be related to (the various derivatives of) a given entanglement measure~\cite{WuLidar04,Campos-Venuti06}.  
It was indeed shown that, given an entanglement measure $M$ related to reduced density operators of the  system, 
first order  phase transition are associated to the anomalies of $M$ while second order phase transitions correspond 
to a singular behavior of the derivatives of $M$. Also quasi-long range order is captured by the behavior of the pairwise 
entanglement\cite{GuLinLi03,Son08}. Other singularities like those  noticed in the concurrence for models with 
three-spin interactions~\cite{Yang05}, are due to the non-analyticity intrinsic in the definition of the concurrence as a 
minimum of two analytic functions and the constant zero. This was then explicitly shown  for the  quantum Ising, 
$XXZ$, and LMG models~\cite{Wu06}.  For the Ising model, for example,  the divergence of the first derivative of the 
concurrence is  determined by the non-analytical behavior of $\langle S^x S^x\rangle$\cite{WuLidar04}.  
A relevant caveat to this approach is constituted by  the uniaxial-LMG  model in a transverse field (with $h^y=0$ 
and $\gamma =0$) that displays a first order QPT for $h^x=0$. The concurrence is continuous at the transition since 
it does not depend on the discontinuous elements of the reduced density matrix~\cite{Vidal-Palacios03}. 
The relation between entanglement and criticality was also studied in  the spin-$1$ $XXZ$ with single ion anisotropy. It 
was established that the critical anomalies  in the entropy experienced at the Haldane-large- $D$ (if  an axial 
anisotropy  $D \sum_i (S^z_i)^2$ is added to the Hamiltonian in Eq.(\ref{haldane-phase})) transition fans out from the 
singularity of the local order parameter $\langle (S^z)^2\rangle$~\cite{Campos-Venuti06}.    
Spontaneous symmetry breaking can influence the entanglement in the ground state.  Below the critical field, the 
concurrence is enhanced by the parity  symmetry breaking~\cite{oster06}. Recently it was demonstrated that such 
enhancement is particularly pronounced for  multipartite entanglement close to the  symmetry breaking.  
This result constitute a further indication  that multipartite, and 
not bipartite, entanglement plays the main role to establish  long-range correlations at the critical points~\cite{oliveira08}. 

In higher dimensions nearly all the results were obtained by means of numerical  simulations. 
The concurrence was computed for the  two dimensional  quantum $XY$ and $XXZ$ models~\cite{Syl2}.
The calculations were based on Quantum Monte Carlo simulations~\cite{QMC-sandvik1,QMC-Syljuasen}.
Although the concurrence for the 2d models results to be qualitatively very similar to the one-dimensional case, it 
is much smaller in magnitude.  The monogamy  limits the entanglement shared among the number of neighbor 
sites. The ground state entanglement in two dimensional XYZ model were analyzed 
in~\cite{Firenze05} by means of quantum Monte Carlo simulations. The divergence 
of the derivative of the concurrence at the continuous phase transition, observed in 
$d=1$, was confirmed; also in this case the range of the pairwise 
entanglement extends only to few lattice sites. By studying 
the one- and the two tangle of the system, it was proved that 
the QPT is characterized by a cusp-minimum in the 
entanglement ratio $\tau_1/\tau_2$. The cusp is ultimately due to the 
discontinuity of the derivative of $\tau_1$. The minimum in the ratio $\tau_1/\tau_2$
signals that the enhanced role of 
the multipartite entanglement in the mechanism driving the phase transition.  
Moreover by looking at the entanglement it was found that 
the ground state can be factorized at certain value of the 
magnetic field. The existence of the factorizing field in $d=2$ was  
proved rigorously for  any $2d-XYZ$ model in a bipartite lattice. 
Unexpectedly enough the relation implying the factorization is very 
similar  to that one found in $d=1$.

\begin{figure}[ht]\centering
\includegraphics[width=8cm]{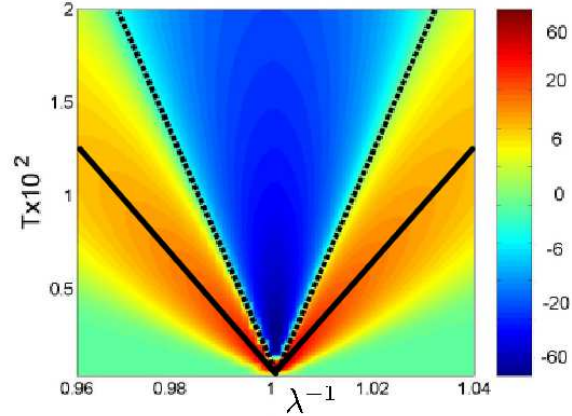}
\caption{The effect of temperature on the anomalies originated from 
the critical divergence of the field-derivative of $C(R)$ can be measured by 
$\partial_T[\partial_{a} C(R)]$. The density plot  corresponds to $\gamma=1$ and 
$R=1$. $T=T^*$ and $T=T_M$  are drawn as dashed and thick lines respectively. 
Maxima below $T^*$ are found  at $T_M =\beta  T_{cross}$ 
with $\beta\sim 0.290 \pm 0.005$ and they are independent of $\gamma$ and $R$; 
the crossover behavior is enclosed in 
between  the two flexes of  $\partial_T[\partial_{a} C(R)]$ at  
 $T_{c1}$ $T_{c2}$; such values are fixed to: $T_{c1}=(0.170  \pm 0.005) T_{cross}$ and 
$T_{c2}=(0.442  \pm 0.005)T_{cross}$ and 
found to be independent of $\gamma$ and $R$. For $T$ smaller than $T_{c1}$ 
$\partial_T[\partial_{a} C(R)]\simeq 0$.
Scaling properties are inherited in $\partial_T[\partial_{a} C(R)]$  
from $\partial_{a} C(R)]$  [From \protect\cite{Amico05}].}
\label{mixedcross}
\end{figure}

\paragraph{Pairwise entanglement at finite temperature}
At finite temperature excitations participate to  entanglement that can become non-monotonous on increasing 
temperature or magnetic field~\cite{Arnesen01,Gunlycke01}. Concurrence for  thermal states, was calculated in several 
situations~\cite{Osborne02,wangth,wangzan,tribedi,Asoudeh04,canosa05,canosa06,rigolinth,zhangzhu,wangwang06,zhangli}.
At finite temperatures but close to a quantum critical points, quantum fluctuations are essential to describe  
the properties of the systems~\cite{Sachdev99}.   For illustration let us consider a one-dimensional quantum $XY$
in an external magnetic field. Although such system cannot exhibits any phase transitions at finite temperature, the very 
existence of  the quantum critical point is reflected in the crossover behavior at $T\ne 0$.
According to the standard nomenclature, the renormalized--classical regime evolves into  the quantum disordered phase
 through the so called
{\em quantum critical region}~\cite{Sachdev99}.  In the $T-h$ plane a $V$-shaped phase diagram emerges, characterized by
the crossover temperature customarily defined as $T_{cross}\doteq |\lambda^{-1}-\lambda_c^{-1}|$.
For $T\ll T_{cross}$ the thermal De Broglie length is much smaller than the average spacing
of the excitations; therefore the correlation functions factorize in two contributions coming from quantum and thermal
fluctuations separately.  The quantum critical region is characterized by $T \gg T_{cross}$.  Here we are in the first regime 
and the correlation functions do not factorize.  In this regime the interplay between quantum and thermal effects is the
dominant phenomenon affecting the physical behavior of the system.
Thermal entanglement close to the critical point of the quantum $XY$ models was recently 
studied by some of us~\cite{Amico05}. In analogy with the zero temperature case it was shown that the  entanglement 
sensitivity to thermal and to quantum fluctuations obeys universal $T\neq 0$--scaling 
laws. The  crossover  to the quantum disordered and renormalized classical 
regimes in the entanglement has been analyzed through the study of derivatives of  
the concurrence $\partial_{\lambda} C$ and  $\partial_T C$. The thermal entanglement 
results to be very rigid when the quantum critical regime is accessed from the 
renormalized classical and quantum disordered regions of the phase diagram; such 
a 'stiffness' is reflected in a maximum in $\partial_T C$ at  $T\sim T_{cross}$. 
The maximum in the derivatives of the concurrence seems a general feature of the 
entanglement in the crossover regime (see for example~\cite{stauber04,Stauber-errata}).
Due to the vanishing of the gap at the quantum critical point,   in the region $T\gg T_{cross}$ an arbitrarily small temperature 
is immediately effective  in the system (see Fig.~\ref{mixedcross}).  
From the analysis  of the quantum mutual information it emerges 
that the contribution of the  classical correlations is negligible in the crossover, thus providing 
the indication that such a phenomenon  is driven solely by the thermal entanglement. 
It is interesting to study how the  existence of the factorizing field $h_f$ affects 
the thermal pairwise entanglement (vanishing at zero temperature). 
It results that the two-tangle $\tau_2$ is still vanishing 
in a region of the $h-T$ plane fanning out from  $h_f$; therefore, if present,   
the  entanglement in the region must be shared between three or more parties. 
In contrast to the analysis 
of the ground state, at finite temperature one cannot characterize the two separate phases of
parallel and antiparallel entanglement.

\paragraph{Thermal entanglement witnesses}
In some case it is hard to quantify the entanglement in a many-body system. Moreover it seems in general 
difficult to relate clear observables to some of the entanglement measures. If one relaxes the 
requirement of quantifying the entanglement and asks only to know if a state is entangled or not then in some 
important cases there is a very appealing answer in terms of the so called entanglement witness.  Interestingly enough it was shown that entanglement witnesses in 
spin systems can be related to thermodynamic quantities~\cite{Toth05a,bruknervedral,wulid05,hide07}. For  the
isotropic $XXX$ or $XX$ Heisenberg model, if the inequality is fulfilled (with $U$ the internal energy, $M^z$
the magnetization)
\begin{equation}
\frac{|U+h^zM^z|}{N|J|} > \frac{1}{4}, 
\label{niko}
\end{equation}
then the  system is in an entangled state.  Once the internal energy and magnetization are 
calculated then it is possible to verify in which range of the parameters of the system and the 
external temperature entanglement is present. Most important is the fact that these types of 
inequalities can be verified experimentally. 
It should be stressed that the analysis based on the entanglement witness 
could be applied to any model for which we can successfully obtain the partition function. 
This feature is the main advantage of using thermodynamic 
witnesses approach to detecting entanglement.  
This method for determining entanglement in solids
within the models of Heisenberg interaction is useful in the cases
where other methods fail due to incomplete knowledge of the
system. This is the case when only the eigenvalues but not
eigenstates of the Hamiltonian are known (which is the most usual
case in solid state physics) and thus no measure of entanglement
can be computed. Furthermore, in the cases where we lack the complete
description of the systems one can approach the problem
experimentally and determine the value of the thermodynamical
entanglement witness by performing appropriate measurements. It is important to 
emphasize that any other thermodynamical function of state could be a 
suitable witness, such as the magnetic susceptibility or heat capacity~\cite{marcin}

\paragraph{Localizable entanglement}
The study of localizable entanglement in spin chains allows to find 
a tighter connection between the scales over which entanglement and correlations decay~\cite{verstraete04,popp05,popp06}.
One expects that the procedure of entangling distant sites by a set of local measurements will be less effective as the 
distance between the two particles increases thus leading to 
a definition of entanglement length $\xi_E$. For a translational invariant system $\xi_E$  can be defined  
in analogy of the standard correlation lenght
\begin{equation}
        \xi_E^{-1} = - \lim_{|i-j| \to \infty} \log \frac{E_{loc}(|i-j|)}{|i-j|} \;.
\label{elenght}
\end{equation} 
By definition the entanglement length cannot be smaller than the correlation length, 
$
        \xi_E \ge \xi
$, 
therefore at a second order phase transition the localizable entanglement length diverges. 
In addition there may also appear "transition points" associated solely to a divergence  
in $\xi_E$. In order to avoid misinterpretations, it must be stressed that
the {\em localizable} ``classical'' two-point correlations then diverge as well. 
For the Ising model in a transverse field it can be shown that~\cite{VerstraetePC04}
$         \max_{\alpha =x,y,z} |Q_{\alpha}^{ij} | \leq
         E_{loc}(i-j)\leq\frac{1}{2}\sum_{\pm}\sqrt{s_{\pm}^{ij}}
$where
$
s_{\pm}^{ij}= \left( 1 \pm \langle S_i^z S_j^z \rangle \right)^2- \left(\langle
S^z_i \rangle \pm \langle S^z_j \rangle \right)^2 \,\, 
$
and 
$
Q_{\alpha}^{ij} = \langle S_i^{\alpha}S_j^{\alpha}\rangle - 
\langle S_i^{\alpha}\rangle \langle S_j^{\alpha}\rangle \;\;\; .
$ 
 In this case, the lower bound  is determined by 
the two-point correlation function in the x-direction.
In the disordered phase ($\lambda < 1$) the ground state possesses a small 
degree of entanglement and consequently its  entanglement length is finite. 
The situation changes at the other side of the critical point. 
Here, although the correlation length is finite, the entanglement length is 
infinite as asymptotically the correlation tends to a finite values. 
The divergence  of $\xi_E$ indicates that the ground state is a globally entangled state,
supporting the general idea that multipartite entanglement is most 
relevant at the critical point~\cite{Osborne02,Firenze04}. 
The properties of localizable entanglement were further investigated for a 
spin-1/2 $XXZ$-chain in~\cite{jin04,popp05} as a function of the anisotropy 
parameter $\Delta$ and of an externally applied magnetic field $h$. 
The authors used exact results for  correlation functions relying on the 
integrability of the  models to find the required bounds.
The presence of the anisotropy further increases the lower bound of the localizable 
entanglement. At the Berezinskii-Kosterlitz-Thouless critical point ($\Delta = 1$) 
the lower bound of the nearest neighbor localizable 
entanglement shows a kink~\cite{popp05}. As pointed out by the authors this might have  
implications in the general understanding of the Berezinskii-Kosterlitz-Thouless 
phase transitions where the ground state energy and its derivatives are continuous as well as the 
concurrence. 
The localizable entanglement in two-dimensional $XXZ$ model was discussed as 
well~\cite{Syl} by means of quantum Monte Carlo simulations. 
A lower bound has been determined by studying the maximum 
correlation function which for $\Delta > -1$ is $Q_{x}$, the long-range 
(power law) decay of the correlation implying a long ranged localizable entanglement.

For half-integer spins, gapped non-degenerate ground states are characteristic for systems in 
a disordered phase (consider paramagnets for example). A finite gap in the 
excitation spectrum of the system in the thermodynamic limit makes the correlations 
decaying exponentially. This is the Lieb-Schultz-Mattis theorem, 
establishing that, under general hypothesis, the ground state of a spin system is 
either unique and gapless or gapped and degenerate~\cite{LIEB} (see \cite{Hastings03} for recent results). It was a surprise, 
when Haldane discovered that systems of integer spins can violate
this theorem~\cite{HALDANE-CONJ1,HALDANE-CONJ2}. 
Accordingly the long range order can be replaced by the so called hidden order of topological nature. 
This kind of order is established in the system because certain solitonic type of excitation become gapless for integer spins\cite{Mikeska}.
This suggests to investigate whether the entanglement in the ground state 
might play some role in establishing the hidden order characteristic for the Haldane phases. 
An aspect that might be relevant to this aim was recently addressed by studying 
the localizable entanglement in  AKLT models~\cite{verstraete04}. 
The ground state of this class of models is of the valence bond type.
For this case it was demonstrated that a singlet state made of two 
spins-$1/2$ located at the ends of the chain can be always realized. This implies that the localizable entanglement is 
long ranged despite the exponentially decaying correlation~\cite{verstraete04}. 
Furthermore the localizable entanglement can be  related to the string 
order parameter. The valence-bond-solid phase order was further studied by looking at the hidden 
order in chains with more complicated topology.  
The von Neumann entropy was studied in spin-$1$ $XXZ$ model with biquadratic 
interaction and single ion anisotropy  in~\cite{GuTianLin05,Wang-spin1} and in~\cite{Campos-Venuti06}.
Some of the features of the corresponding phase diagram are captured.
The Haldane transitions exhibited in the phase diagrams are marked by 
anomalies  in the Von Neumann entropy; its  maximum at the 
isotropic point is not related to any critical phenomenon 
(the system is gapped around such a point), but it is due to the 
equi-probability of the three spin-$1$ states occurring at that point~\cite{Campos-Venuti06}. 
Since the  Berezinskii-Kosterlitz-Thouless 
transition separating the $XY$ from the Haldane or large-$D$ phases connects
a gapless with a gapped regime, it was speculated that an anomaly in the entanglement 
should highlight such transition~\cite{GuTianLin05}.

\begin{figure}\begin{center}
\includegraphics[scale=0.3,angle=0]{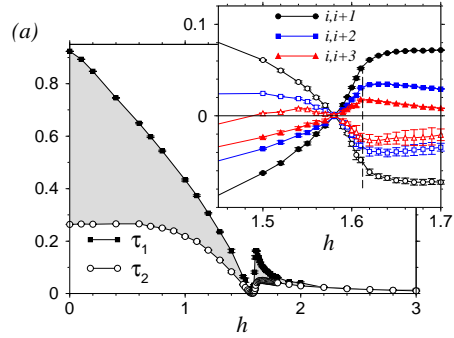}\end{center}
\caption{One-tangle $\tau_1$ and the sum of squared concurrences 
$\tau_2$ as a function of the applied magnetic field in $z$-direction
for the $XYZ$ model with exchange along $y$: $J_y= 0.25$ (in unit of exchange along $z$). 
Inset: contributions to the concurrence between j-th neighbors; full
symbols stand for $C^{I}(j)$, open symbols for $C^{II}(j)$. The dashed
line marks the critical field $h_c$.  [From \protect\cite{Firenze04}] } 
\label{tau2tau1}
\end{figure}

\paragraph{Multipartite entanglement}
Although pairwise entanglement allow to capture important properties of the phase diagram it was evidenced 
that the  spin systems are most generically in multipartite entangled state\cite{wang02,stelmachovic04,bruss05}.
Just to make an example we point out that the first excited state above a ferromagnetic ground state 
is a W-state that is a well-known state in quantum information with a multipartite entanglement.
Despite its importance, a   {\it quantitative}  description of multipartite entanglement constitutes a challenging problem in 
the current research. In many-body physics multipartite entanglement 
has been studied resorting to 'global' measures that most often cannot  distinguish 
different types of multipartite entanglement each other (see however \cite{guehne05}). 
A first way to estimate multipartite entanglement in spin system is provided 
by the entanglement ratio $\tau_2/\tau_1$ as the amount of two spin  
relative to global entanglement. 
It is interesting to compare the behavior of such quantitities for quantum critical and factorizing points 
of spin models.  
In fact it emerged that  $\tau_2/\tau_1$ is small close to quantum critical points. In contrast the entanglement ratio 
approach  to one close to the factorizing point. 
Close to quantum critical points the entanglement ratio was calculated numerically for  
$1d-XYZ$\cite{Firenze04}. 
(Fig.\ref{tau2tau1}).
The entanglement ratio was calculated close to factorizing points for the quantum Ising 
model.

To address multipartite entanglement directly several routes have been suggested. 
Of interest is the analysis based on geometric entanglement. Most of the works till now concentrated on critical systems.
It was proved that geometric entanglement obeys an area law that, most probably, coincides 
with the well established area law of the von Neumann entropy~\cite{botero07,orus08,vidal08,orus.pra08,orus09}.Such a result 
provides a further evidence  that the universal behaviour of the  
block entanglement close to a critical point traces back to 
multipartite entanglement.

Although some of the  proposed measures rely on  $n$-point correlation functions there is no clear evidence on whether this is a general feature of multipartite entanglement. 
As important example of measure of multipartite entanglement relying on two-point function
was proposed in~\cite{oliveira,oliveira06b,Somma04} as 
\begin{equation}
    E^{(2)}_{gl} = \frac{4}{3}\frac{1}{N-1}\sum_{l=1}^{N-1} \left[ 
    1 - \frac{1}{N-1} \sum_{j=1}^N \mbox{Tr} \rho_{j,j+l}^2 \right]
\label{E2}
\end{equation}
where $\rho_{j,j+l}$ is the reduced density matrix associated to the sites $j$ and 
$j+l$.
 Similarly one can consider also three-body reduced density matrices and 
construct the corresponding global entanglement measure. Although the precise form  has not been established yet, the 
global entanglement $ E^{(n)}_{gl}$, generalization of Eq.(\ref{E2}) should be related to the
set of reduced n-qubits density operators. According to Oliveira and coworkers, then  the hierarchy of $E^{(n)}$  might  provide 
a comprehensive description of entanglement in many-body systems already for moderate values of $n$. 

Global  entanglement 
is very sensitive to the existence of QPTs. 
As a paradigmatic example the authors analyzed the phase diagram in the anisotropy-magnetic  field plane. 
By extending an earlier approach developed in~\cite{WuLidar04}, 
de Oliveira {\em et al.} also showed how the non-analytic behavior of $E^{(n)}_{gl}$ 
is related to that of the ground state energy. Note that from Eq.(\ref{E2}) it is 
possible to define an entanglement length proportional to the correlation length $\xi$. 
This differs considerably from that one defined by the localizable 
entanglement (see Eq.(\ref{elenght})); the latter is always bounded from below by the 
correlation length and can even be divergent where $\xi$ is finite.

As discussed in~\cite{facchi06,facchi06a,facchi06b} the analysis of the average purity 
might not be sufficient and the analysis of the distribution of the purity for 
different partitions could give additional information. 
Rather than measuring multipartite entanglement in terms of a single number, one 
characterizes it by using a whole function. One studies the distribution function 
of the purity (or other measures of entanglement) over all bipartitions of the
system. If the distribution is sufficiently regular, its average and
variance will constitute characteristic features of the global entanglement: the
average will determine the ``amount'' of global entanglement in the
system, while the variance will measure how such entanglement
is distributed. A smaller variance will correspond to a larger
insensitivity to the choice of the bipartition and, therefore, will
be characteristic for different types of multipartite entanglement. 

In \cite{patane.bound} multipartite entanglement is studied with the aim to shed light on how entanglement is shared in a many-body system. For a quantum $XY$ model the simplest 
multiparticle entanglement   of a subsystem made of three arbitrary spins of the chain is considered; then bipartite entanglement 
between a spin and the other two with respect to all possible bipartitions. It is found  that  the block of two spins  may be  entangled with the  external spin, 
despite the latter  is not entangled directly with any of the two spins separately 
(see Fig. \ref{NegT=0}). 
Hence the range of such spin/block entanglement  may extend further than  
the spin-spin entanglement range.
It is plausible that increasing the size of the subsystem  considered
will increase the range of the multipartite entanglement.
For instance, the range of  spin/block entanglement 
  will increase if we consider a larger block.
Hence, a  single spin  can be  entangled  with more distant partners,
if one allows to cluster them into a large enough block.
It would be  intriguing  to study how   spin/block entanglement and,
 in general, block/block entanglement between subsystems,  scale increasing the size of blocks.
especially exploring the  connection with quantum criticality.
We remark that such analysis would be different
with respect to the well known block entropy setting,
since in that case one is interested in the block/rest-of-the-system entanglement.

\begin{figure}[ht]\centering
\includegraphics[width=5cm]{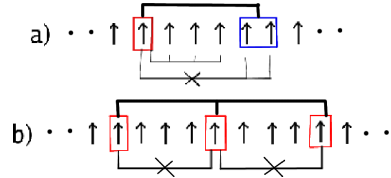}
\caption{ Configurations of spins described  in the text and whose entanglement properties 
are  presented in Fig. \ref{NegT=0}.
 We fix $R=3$, hence a spin is directly entangled with its first three
nearest neighbors.
 {\bf a)}  'Clustering' two spins increases the range of entanglement 
(the scheme is symmetric also for spins on the left of the  marked one). 
{\bf b)} Symmetric configuration of spins such that no two-particle entanglement is present,  
 but still the spins share multiparticle entanglement.} 
\label{clustering}
\end{figure}

\begin{figure}[ht]
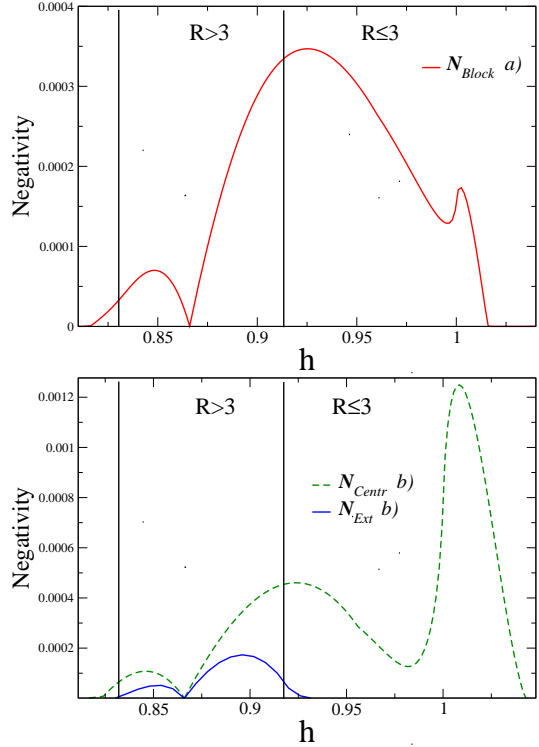
\centering
\includegraphics[width=7cm]{figures/Nega.eps}
\includegraphics[width=7cm]{figures/Negb.eps}
\caption{$T=0$ Negativities between one spin and the other two Vs magnetic field $h$ are shown
for both configurations of Fig \ref{clustering}.
 We consider $\gamma=0.5$. In this case outside the interval
 marked by the two solid vertical lines 
the range of spin-spin entanglement is $R\leq 3$
(for values of $h$ inside this interval  $R$ grows
due to the its divergence at factorizing field $h_f=\sqrt{1-\gamma^2}\simeq 0.86$ \cite{Amico06}).
For configuration $a)$ (upper panel), $\mathcal{N}_{Block}$  signals the spin/block entanglement for a distance $d=4$.
For values of $h$ outside  the vertical lines, the Negativity signals genuine spin/block entanglement.
For configuration $b) $ (lower panel), both  $\mathcal{N}_{Ext}$ (solid line) and  $\mathcal{N}_{Centr}$ (dashed line) are plotted.
For values of $h$  outside the vertical lines the three spins share no spin/spin entanglement,
hence  for  non zero $\mathcal{N}_{Ext}$ and $\mathcal{N}_{Centr}$
 free multiparticle entanglement is present.
 The latter turns in to bound entanglement for values of  $h$ such that only
$\mathcal{N}_{Centr}\ne 0$ and  $\mathcal{N}_{Ext}=0$ (both on the left and on the right of the solid lines).  [From\cite{patane.bound}]}
\label{NegT=0}
\end{figure}

\paragraph{Bound entanglement}
We would like to conclude this brief description of the relation between entanglement and magnetic order by 
analyzing in which cases interacting spin system are in a  bound entangled state (see section (\ref{meas}). We 
shall see that such kind of peculiar entangled states are  generically 'engineered' by a many-body system at equilibrium
as it   occurs naturally in certain region of the phase diagram
for the 'last' entangled states before the complete separability is
reached. In this sense the bound entanglement bridges between 
quantum and classical correlations. Being the bound entanglement 
a form of demoted entanglement, it appears when quantum correlations get weaker.
Bound entangled states were found in both the ground and thermal states of anisotropic $XY$ models\cite{patane.bound,acin.bound1,acin.bound2}. 
We follow  the approach pursued in \cite{patane.bound}  where 
 three-spin entanglement in an infinite anisotropic $XY$ chain was analyzed).
Two different configurations were considered (see Fig.\ref{clustering}).

\begin{figure}[ht]\centering
\includegraphics[width=8cm]{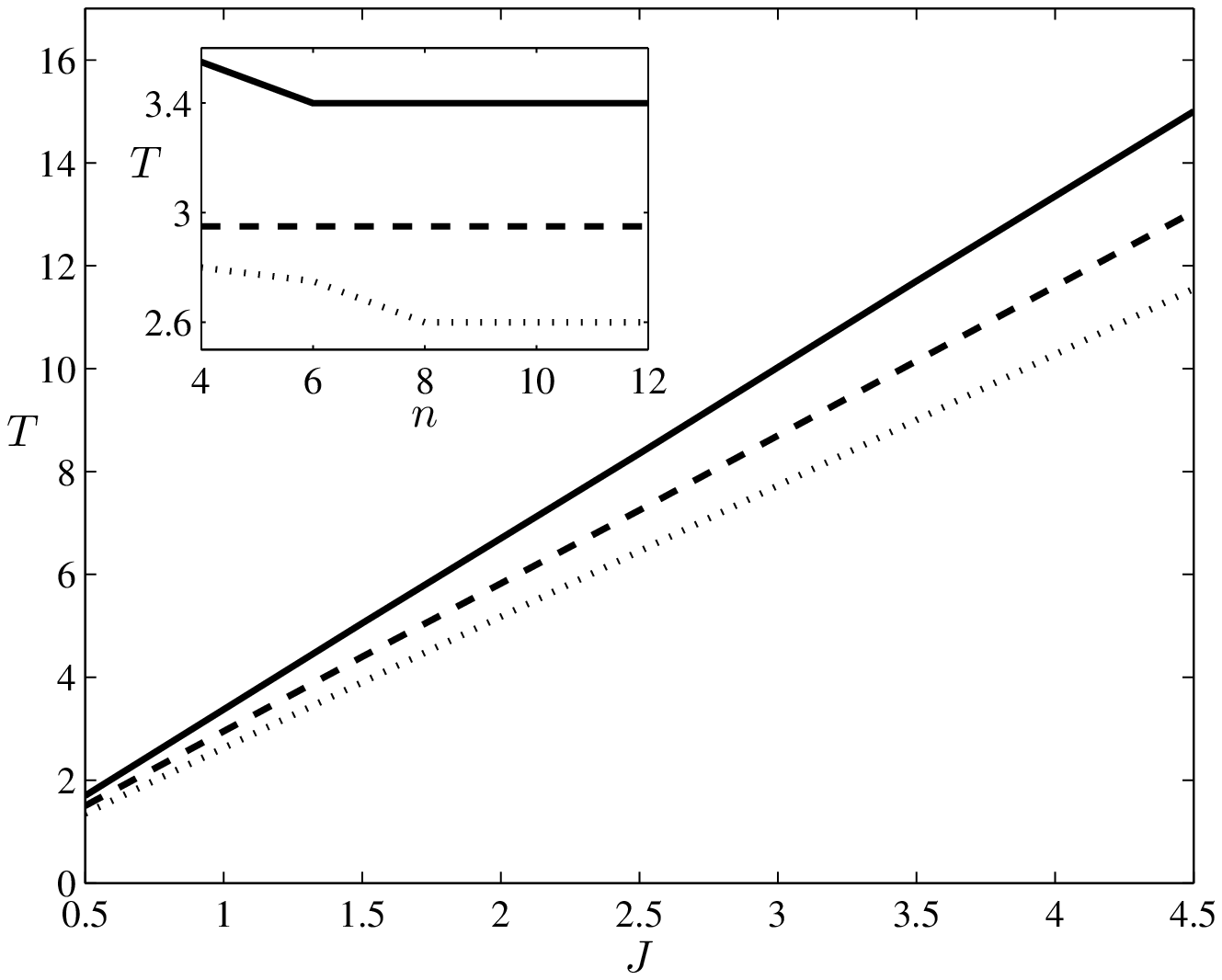}
\caption{
Threshold temperatures above that the negativity is
    zero in the even-odd (solid line), $1:n-1$
    (dotted line) and
    half-half (dashed line) partitions. We plot the threshold
    temperature as a function of the coupling parameter $J$ of the model (\ref{general-spin})
with
$J^{(ij)}_x=J^{(ij)}_y=\delta_{i+1,j}$ and $J^{(ij)}=0\, \forall ij$
    for  $n=10$ spins and $h_z=1.9$.
    {\bf{Inset:}} Temperature above which the negativity in the
    even-odd (solid line), $1:n-1$
    (dotted line) and half-half (dashed line) partitions is zero as a function of the
    number $n$ of the spin  Hamiltonian (\ref{general-spin}) with $J=1$ and $h_z=1.9$.[From\cite{acin.bound2}]}
\label{thermal}
\end{figure}

At zero
temperature bound entanglement appears  (see Fig. \ref{NegT=0}) when
the spins are sufficiently distant each others  and as in the case
of the spin/spin entanglement, it can be arbitrary long ranged near
the factorizing field. 
To prove it, the idea is to resort  the "incomplete separability" condition described in the first section.
In fact from  Fig. \ref{NegT=0} we see that  $\mathcal{N}_{Ext}$  may be zero even if  $\mathcal{N}_{Centr}$ is non-zero.
 Thus in such case the density matrix of the spins is  PPT for the two symmetric bipartitions of one external spin vs the other two
($\uparrow|\uparrow \uparrow$ and $\uparrow\uparrow |\uparrow$) and
 Negative Partial Transpose  (NPT) for the partition of the central spin vs the other two (we remark that PPT does not 
 ensure the separability of the two partitions for dimensions of local Hilbert space greater than two).
In fact if we could be able to distill a maximally entangled state between two spins
  then one of two previous PPT partitions would be NPT
and this cannot occur since PPT is invariant under LOCC \cite{Horo-bound,Vidal02}.
Quantum states must be  'mixed enough'  to be bound entangled.   
In the ground state  a source mixing is the 
trace over the other spins of the chain. However  if the spins
are near enough the reduced entanglement is free. It results that  
the effect of the thermal mixing can drive the $T=0$  free entanglement to bound entanglement (see Fig. \ref{thermal}). 
This behavior shown for the Ising model is also found  for the
entire class of quantum XY Hamiltonians with generic values of anisotropy (the 
temperature at which the different types of entanglement  are
decrease with $\gamma$.
NPT thermal bound entanglement was also found by \cite{acin.bound1,acin.bound2}
resorting to block entropies. The idea of Acin and coworkers was to calculate the block entropy for two different bipartitions: 
one in which the two  subsystems are made of contiguous block of spins  (called 'half-half' partition); the other groups 
all the  spins labeled, say, by even 
indices in one subsystem and the remaining ones in the other (called 'even-odd' partition). 
Because of the area law, the entanglement in the even-odd partition is more robust to thermal fluctuations than 
entanglement in the half-half partition 
(an hypothesis corroborated by actual calculations by the same authors). Therefore there is a range of 
temperatures for which  the PPT condition is reached with an even-odd entanglement. The entanglement 
is bound because single particles cannot distill entanglement (as the half-half bipartion can be singled 
out to have particles in two different  blocks).  The calculations are  done for a finite set of spins interacting 
according to an isotropic $XY$ Hamiltonian.
We remark that such NPT bound entanglement was found in closed systems. 
Recently it was demonstrated that NPT bound entanglement can arise dynamically from decoherence of multipartite entangled 
states of GHZ type\cite{acin.bound.deco}. Based on that it is intriguing to conjecture that such kind of entanglement could be 
generated dynamically via decoherence in open systems.
Besides NPT, also PPT bound entangled states were found in spin systems at finite temperature\cite{toth.bound1,toth.bound2}.  
The method developed by Toth {\it et. al}  relies on certain relation between entanglement and squeezing of collective spins 
that can serve as separability test for separability of the given state (like an entanglement  witness). The entanglement detected 
can be of multipartite type despite the relations involve only two-point correlation functions. They considered spin models of the 
Heisenberg type at finite size and proved that a range of temperature exists where their thermal state display multipartite 
entanglement that cannot be distilled for any bipartition of the system.

\section{Conclusions and outlook}
\label{concl}

The use of concepts developed in quantum information science has provided a new twist to 
the study of many-body systems. Here we presented a specific example of this kind of 
approach by discussing the relation between magnetism and entanglement. Looking at 
the next future it seems to us that the most challenging problems are a wider characterization 
of the multipartite entanglement and, in our opinion most important, a connection between 
this acquired knowledge and new experiments. Remarkable  impact 
of quantum information in condensed matter has been proving on the possibility to design more 
efficient classical numerical algorithms for quantum many-body systems.

As for  experimental tests on  entanglement 
in many-body systems, we observe that 
the most direct method seems relying  on the entanglement witnesses that 
have been derived using thermodynamical quantities. Nevertheless methods 
based on neutron scattering techniques on magnetic compounds,   
that are of particular  relevance for spin systems, are also valuable especially      
for a direct quantification of entanglement in macroscopic  systems. In this context  we notice that more refined 
experimental analysis seem to be required to extract entanglement.
These  might disclose  new unexplored  features of entangled many-body states.  

\ack
We are very grateful to  A. Osterloh and V. Vedral for numerous discussions on the topic 
covered by this article. We thank  F. Baroni, P. Calabrese, G. De Chiara, A. Fubini, V. Giovannetti, S. Montangero,  A. Osterloh, 
G.M. Palma, D. Patan\`e, F. Plastina,  D. Rossini, W. Son, V. Tognetti, V. Vedral, and P. Verrucchi  for fruitful collaboration. 
This work was supported by IP-Eurosqip, SNS-Research Project and National Research Foundation \& Ministry of Education, Singapore
\bigskip

\section*{References}
\bibliographystyle{jphysicsB}

\end{document}